\documentclass{aa}  
\usepackage{graphicx}
\usepackage{txfonts}
\begin{document}
   \title{Upper limit for circumstellar gas around the Type Ia
   SN~2000cx}

   \author{F. Patat
          \inst{1},
          S. Benetti\inst{2},
          S. Justham\inst{3},
          P.A. Mazzali\inst{4,5},
          L. Pasquini\inst{1},    
          E. Cappellaro\inst{2},\\
          M. Della Valle\inst{6},    
          Ph.\ Podsiadlowski\inst{3},
          M. Turatto\inst{2},
          A. Gal-Yam\inst{7}
          \and
          J.D. Simon\inst{7}\fnmsep\thanks{Based on observations obtained
	  at ESO-Paranal.}
          }

   \offprints{F. Patat}

   \institute{European Organization for Astronomical Research in the
	Southern Hemisphere; K. Schwarzschild-Str. 2, 85748 Garching, Germany\\
              \email{fpatat@eso.org}
         \and
             INAF - Osservatorio Astronomico di Padova, v. Osservatorio n.5,
	35122 Padova, Italy
         \and
           Department of Astrophysics, University of Oxford, Oxford OX1 3RH, UK
         \and
             Max-Planck-Institut fuer Astrophysik, K. Schwarzschild Str.2,
           85748, Garching, Germany
         \and
            INAF - Osservatorio Astronomico di Trieste, v. Tiepolo 11,
            34131 Trieste, Italy
         \and	
           INAF - Osservatorio Astronomico di Arcetri, Largo E. Fermi n.5,
	Firenze, Italy
         \and
         Astronomy Department, MS 105-24, California Institute of
         Technology, Pasadena, CA 91125,USA\\
             }

   \date{Received August 1, 2007; accepted August 27, 2007}

 
  \abstract
   {The nature of the companion stars in Type Ia Supernova (SNe) progenitor 
   systems remains unclear. One possible way to discriminate between different
   scenarios is the presence (or absence) of circumstellar material, the
   left overs from the progenitor evolution that may be revealed by their
   interaction with the SN.}
   {A new method to probe the circumstellar environment 
   has been exploited for the normal Type Ia SN~2006X, leading for the first 
   time to the direct detection of material which escaped the progenitor 
   system. In this paper we apply the same analysis to the peculiar 
   Type Ia SN~2000cx, with the aim of constraining the properties of its 
   progenitor system.}
   {Using multi-epoch, high-resolution spectroscopy we have studied the 
   spectral region where narrow, time-variable Na~ID absorption features are 
   expected in case circumstellar material is present along the line of 
   sight.}
   {No Na~ID absorption is detected in the rest-frame of the host galaxy
   to a level of a few m\AA, setting a stringent upper limit to the column 
   density of the absorbing material 
   ($N(NaI)\leq $2$\times$10$^{10}$ cm$^{-2}$).} 
   {In this respect the peculiar Type Ia SN~2000cx is different 
    from the normal Ia SN~2006X. Whether this 
    is to be attributed to a different progenitor system, to viewing-angle 
    effects or to a low metallicity remains to be clarified.}


\authorrunning{F. Patat et al.}

   \maketitle
%

\section{Introduction}
\label{sec:intro}

Due to their enormous luminosities and their homogeneity, Type Ia
Supernovae (hereafter SNe Ia) have been used in cosmology as standard
candles, with the ambitious aim of tracing the evolution of the
Universe (Riess et al. \cite{riess}; Perlmutter et
al. \cite{perlmutter}). Despite of the progresses made in this field,
the nature of the progenitor stars and the physics which governs these
powerful explosions are still uncertain.  In general, they are thought
to originate from a close binary system (Whelan \& Iben
\cite{whelan}), where a white dwarf accretes material from a companion
until it approaches the Chandrasekhar limit and finally undergoes a
thermonuclear explosion. This scenario is widely accepted, but the
nature of both the accreting and the donor star is not yet known, even
though favorite configurations do exist (Branch et al. \cite{branch};
Parthasarathy et al. \cite{partha}; Tutukov \& Fedorova
\cite{tutukov}). A discriminant between some of the proposed scenarios
would be the detection of circumstellar material (CSM).  However,
notwithstanding the importance of the quest, all attempts of detecting
direct signatures of the material being transferred to the accreting
white dwarf in normal SNe~Ia were so far frustrated, and only upper
limits to the mass-loss rate could be placed from optical (Mattila et
al.\  \cite{mattila}), radio (Panagia et al.\  \cite{panagia06}) and
UV/X-Ray emission (Immler et al. \cite{immler}).  Claims of possible
ejecta-CSM interaction have been made for a few normal objects, namely
SN~1999ee (Mazzali et al.  \cite{mazzali05}), SN~2001el (Wang et al.
\cite{wang03}), SN~2003du (Gerardy et al.\  \cite{gerardy}) and
SN~2005cg (Quimby et al. \cite{quimby06}).  In all those cases, the
presence of CSM is inferred from the detection of high-velocity
components in the SN spectra. However, it must be noticed that these
features can be explained by a 3D structure of the explosion (Mazzali
et al. \cite{mazzali05}) and, therefore, circumstellar interaction is
not necessarily a unique interpretation (Quimby et
al. \cite{quimby06}). Furthermore, even if the high-velocity
components were indeed caused by an ejecta-CSM interaction, it would
not be possible to estimate the velocity or density of the CSM.

Two remarkable exceptions are represented by the peculiar SN~2002ic
and SN~2005gj, which have shown extremely pronounced hydrogen emission
lines (Hamuy et al. \cite{hamuy}; Aldering et al. \cite{aldering};
Prieto et al. \cite{prieto}), that have been interpreted as a sign of
strong ejecta-CSM interaction.  However, the classification of these
supernovae as SNe Ia has been questioned (Benetti et
al. \cite{benetti06}), and even if they were SN Ia, they must be rare
and hence unlikely to account for normal Type Ia explosions (Panagia
et al. \cite{panagia06}). As a matter of fact, the only genuine
detection may be represented by the underluminous SN~2005ke (Patat et
al. \cite{patat05}), which has shown unprecedented X-ray emission
at the 3.6$\sigma$-level accompanied by a large UV excess (Immler et
al. \cite{immler}). These observations have been interpreted as the signature
of a possible weak interaction between the SN ejecta and material lost
by a companion star (Immler et al. \cite{immler}). Interestingly, the
SN has not been detected at radio wavelengths (Soderberg
\cite{soderberg}).

A different approach has been recently proposed and tested on the
normal SN Ia 2006X (Patat et al. \cite{patat}; hereafter P07). For
SN~2006X, P07 detected material within a few 10$^{16}$ cm from the
explosion site. Based on the velocity, density and location of the
CSM, P07 have concluded that the favorite companion star for the
progenitor of SN~2006X is a red giant at the time of explosion. The
method is based on the study of time variation of absorption lines
(Na~I, Ca~II, K~I) arising in the CSM. Since in SNe Ia the UV
bluewards of 3500\AA\/ undergoes severe line blocking by heavy
elements like Fe, Co, Ti and Cr (Pauldrach et al. \cite{pauldrach};
Mazzali \cite{mazzali00}), they are able to ionize any existing
circumstellar material only within a rather small radius. Once the SN
UV flux has significantly decreased in the post-maximum phase, the
material that has a sufficiently high density recombines, producing
time-variable absorption features. In particular, due to their
strength, the high-resolution study of the evolution of Na~I D lines
offers a powerful diagnostic, capable of revealing very small amounts
of material along the line of sight, without requiring direct
interaction between the SN ejecta and CSM.

In order to investigate the existence of multiple channels to Type Ia
explosions and the possible presence of viewing angle effects (which
are expected if the material is confined within a disk or a torus), a
large sample of Type Ia events needs to be studied. The drawback of
the novel approach proposed by P07 is that it requires multi-epoch,
high-resolution ($\Delta\lambda/\lambda \geq$40,000) and high
signal-to-noise ratio ($SNR\geq$50) data, making it feasible only with
a large amount of integration time at large telescopes and for
close-by objects only. This is also one of the reasons that make the
data set of SN~2006X unique, at least in this respect.

A few Type Ia SNe have been observed at high spectral resolution with
the aim of detecting possible signs of interaction with CSM (Lundqvist
et al. \cite{lundqvist03, lundqvist05}; Sollerman et
al. \cite{sollerman05}; Mattila et al. \cite{mattila}). In particular,
searching the ESO-VLT Archive\footnote{http://archive.eso.org/}, we
have found one object, namely the peculiar Type Ia SN~2000cx (Li et
al. \cite{li}), that has been observed with VLT-UVES by two different
groups, covering two pre-maximum epochs (discussed by Lundqvist et
al. \cite{lundqvist03}) and two additional epochs at about two months
past maximum (unpublished, see next section). SN~2000cx was discovered
about ten days before maximum in the S0 galaxy NGC~524 (Yu, Modjaz \&
Li \cite{yu}) and was classified as Type Ia based on its similarity to
SN~1991T (Chornock et al. \cite{chornock}). SN~2000cx may be as
luminous as SN~1991T (see for example Mazzali et al. \cite{zorro}),
but it does not obey to the luminosity vs. light curve width relation
(Li et al. \cite{li}) and it is less peculiar from a spectroscopic
point of view, since it does not show a Fe lines dominated spectrum at
the earliest phases (Filippenko et al. \cite{flipper92}; Ruiz-Lapuente
et al. \cite{lapolenta}; Mazzali, Danziger \& Turatto
\cite{mazzali95}).

Multi-epoch photometry, low-resolution optical and near IR
spectroscopy for SN~2000cx have been presented by Li et
al. (\cite{li}), Rudy et al. (\cite{rudy}), Sollerman et
al. (\cite{sollerman04}) and Branch et al. (\cite{branch04}). Here we
focus on the analysis of the high-resolution spectroscopic data
obtained for SN~2000cx on four epochs, ranging from day $-$7 to day
+68 with respect to $B$ maximum light.

The paper is organized as follows. In Sec.~\ref{sec:obs} we describe
the observations and the data reduction, while in
Sec.~\ref{sec:ngc524} we illustrate the properties of the host
galaxy. The results are presented in Sec.~\ref{sec:hires} and
discussed in Sec.~\ref{sec:disc}. Finally, in Sec.~\ref{sec:concl} we
summarize our conclusions.

\section{Observations and Data Reduction}
\label{sec:obs}

The high-resolution data were obtained with the ESO 8.2m Kueyen
Telescope equipped with the Ultraviolet and Visual Echelle
Spectrograph (UVES; Dekker et al. \cite{dekker}) on several epochs,
using different instrumental setups (DIC1 390+564, DIC2 437+860, RED
580).  For our purposes we have selected DIC1 390+564 and RED 580,
retaining only the data sets obtained on 2000-07-20, 2000-07-25,
2000-09-17 and 2000-10-03. The other epochs (2000-09-12, 2000-09-13
and 2000-09-14) were affected by bad weather and did not add
significantly to the phase coverage.  The main information on the data
is recapped in Table~\ref{tab:obs}.

\begin{table}
\centerline{
\tabcolsep 1.7mm
\begin{tabular}{cccccc}
\hline
Date & Phase & Setup & Range & Exp. Time & Slit Width\\
(UT) & (days)&       & (\AA) & (seconds) & (arcsec)\\
\hline
           &      &      &           &               &   \\
2000-07-20 & $-$7 & DIC1 & 3280-4510 & 2$\times$2400 & 0.8\\
           &      &      & 4620-5600 & \\
           &      &      & 5675-6650 & \\
2000-07-25 & $-$2 & DIC1 &     -     & 3$\times$3600 & 0.8 \\
2000-09-17 &  +52 & RED  & 4780-5760 & 2$\times$5400 & 1.0 \\
           &      &      & 5835-6810 &               & \\
2000-10-03 &  +68 & RED  &     -     & 2$\times$5400 & 1.0 \\
\hline
\end{tabular}
}
\caption{\label{tab:obs} VLT-UVES observations of SN~2000cx used in this
paper. Phase refers to $B$-band maximum light (July 26.7, 2000, Li et
al. \cite{li}).}
\end{table}

The data were reduced using the UVES Data Reduction Pipeline
(Ballester et al. \cite{ballester}). Wavelength calibration was
achieved using Th-Ar lamps, with a final RMS accuracy of 0.15
km s$^{-1}$. The wavelength scale was corrected to the rest-frame
adopting a host galaxy heliocentric recession velocity of 2353 km
s$^{-1}$ (Emsellem et al. \cite{emsellem}). To compensate for the
Earth's motion, a heliocentric velocity correction was applied to
the data. The atmospheric lines were identified using a
spectroscopically featureless bright star (HR~3239) observed with the
same instrumental setup as for the science data. The resolving power
$\Delta\lambda/\lambda$ at 5900\AA\/ is 54,000 (DIC1) and
42,600 (RED), corresponding to a full width half maximum (FWHM)
resolution of 5.5 and 7.0 km s$^{-1}$ for the two setups,
respectively. Finally, in order to increase the signal-to-noise ratio,
spectra obtained on the same night were combined. The resulting
signal-to-noise ratios per pixel (0.017\AA) in the Na~I D region are
90, 160, 70 and 50 for day $-7$, $-$2, +52 and +68, respectively.

The gas column densities relative to the detected absorptions were
estimated using VPGUESS\footnote{ VPGUESS has been developed by
J. Liske and can be freely downloaded at
http://www.eso.org/$\sim$jliske/vpguess/index.html} and
VPFIT\footnote{VPFIT has been developed by R.F. Carswell and can be
freely downloaded at http://www.ast.cam.ac.uk/$\sim$rfc/vpfit.html}.

\begin{figure}
\centerline{
\includegraphics[width=9cm]{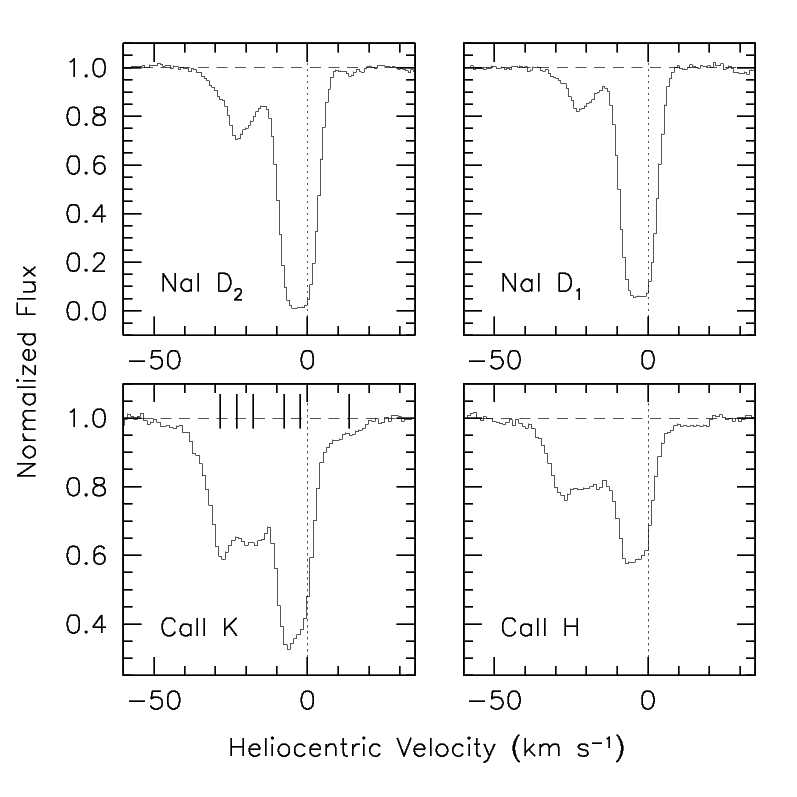}
}
\caption{\label{fig:mw} Line profiles for the Galactic Ca~II H\&K 
and NaI~D lines on day $-$2. The resolution is 5.5 km s$^{-1}$ (FWHM).
the vertical tickmarks on the Ca~II K diagram indicate the positions
of the main components.}
\end{figure}

\section{Host galaxy properties and SN location}
\label{sec:ngc524}

NGC~524 is classified as S0$^+$(rs) (de Vaucouleurs et
al. \cite{devauc}), and it shows a gaseous disk with some hint of
spiral pattern in the nuclear region (Sarzi et al. \cite{sarzi}).
Considering its low ellipticity ($\epsilon<$0.05, Simien \& Prugniel
\cite{simien}), it shows very high velocities, with an amplitude that
reaches $\sim$140 km s$^{-1}$ at $r \simeq 20$\,\arcsec (Emsellem et
al. \cite{emsellem}). The metallicity index Mg\,$b$\/ presents regular
gradients as in the case of the H$\beta$ map, which shows a mild
positive gradient (Kuntschner et al. \cite{kunt}).  The metallicity in
the central part of the galaxy is estimated to be [Z/H]$\sim$0.15,
while it decreases by about $-$0.35 per dex in $\log (r/r_{eff})$
(H. Kuntschner, private communication).

SN~2000cx exploded at a projected distance of 111.7 arcsec from the
nucleus of NGC~524 (Yu et al. \cite{yu}), corresponding to $\sim$2.3
effective radii (Kuntschner et al. \cite{kunt}). At a distance of 33
Mpc ($H_0$=72 km s$^{-1}$ Mpc$^{-1}$), this corresponds to about 18
kpc. The velocity maps of NGC~524 (Emsellem et al. \cite{emsellem})
extend to about 35$^{\prime\prime}$ from the nucleus in the SW
direction (where they reach a velocity of about $-$140 km s$^{-1}$)
and therefore do not include the apparent position of the explosion
site.  Nevertheless, these maps suggest that the SN is projected onto
the approaching side of the host galaxy. Preliminary results from GMOS
observations of NGC524 reaching out to about 2.5 times the effective
radius are consistent with the extrapolation of the SAURON velocity
maps, albeit indicating velocities of only $\sim-$100 km s$^{-1}$ at
the position of SN~2000cx (M. Norris, private communication).

\section{Absorption features}
\label{sec:hires}

\subsection{Milky Way}
\label{sec:mw}

The Milky Way reddening along the line of sight to NGC~524 is
$A_B$=0.356 ($E(B-V)$=0.083, Schlegel, Finkbeiner \& Davis
\cite{schlegel}) so that  narrow absorption features are expected to
be detectable. Indeed, the UVES data we present here clearly show
strong Ca~II H\&K and Na~I D lines at heliocentric velocities of 0-30
km s$^{-1}$ (see Fig.~\ref{fig:mw}). The global equivalent
widths ($EW$) of these lines are 0.24$\pm$0.01 \AA\/ (K),
0.14$\pm$0.01 \AA\/ (H), 0.35$\pm$0.01 \AA\/ (D$_2$) and 0.29$\pm$0.01
\AA\/ (D$_1$), respectively. These features show no significant 
variation during the epochs covered by our data (see
Table~\ref{tab:ew}). The ratio of the color excess and the total
equivalent width of Na~I D lines is $E(B-V)/EW(NaI D)$=0.13, placing
SN~2000cx on the lower slope relation between $EW$ and $E(B-V)$
presented by Turatto, Benetti
\& Cappellaro (\cite{turatto}).

\begin{table}
\centerline{
\begin{tabular}{ccccc}
\hline
Epoch & Ca~II K & Ca~II H & Na~I D$_2$ & Na~I D$_1$\\
(days)& (\AA)   & (\AA)   & (\AA)      & (\AA)\\
\hline
$-$7  & 0.24$\pm$0.02 & 0.14$\pm$0.02 & 0.36$\pm$0.03 & 0.29$\pm$0.03\\
$-$2  & 0.24$\pm$0.01 & 0.14$\pm$0.01 & 0.35$\pm$0.01 & 0.29$\pm$0.01\\
+52   &   -           & -             & 0.35$\pm$0.02 & 0.29$\pm$0.02\\
+68   &   -           & -             & 0.36$\pm$0.03 & 0.29$\pm$0.03\\
\hline
\end{tabular}
}
\caption{\label{tab:ew} Equivalent widths of Ca~II and Na~I Galactic 
features.}
\end{table}

Closer inspection of the line profiles shows a number of
sub-components, better seen in the less saturated Ca~II H\&K
lines. The main features correspond to +13.5, $-$2.4, $-$7.5, $-$17.7,
$-$23.1 and $-$28.5 km s$^{-1}$ (see Fig.~\ref{fig:mw}), the most
intense being the second and the third component. The total column density
deduced from the Voigt profile fitting of the D$_1$ component gives
$\log N(NaI)$=13.1$\pm$0.1. For a typical Milky Way dust mixture this
would imply $E(B-V)$=0.24$\pm$0.03 (Hobbs
\cite{hobbs}), which is almost a factor 3 larger than the value
reported by Schlegel et al.\ (\cite{schlegel}). Even though the column
density estimate is certainly hampered by line saturation
($EW(D_2)/EW(D_1)\sim$1.2), we notice that the column density
corresponding to $E(B-V)$=0.083 is $\log N(NaI)$=12.3, which can
hardly account for the observed line depth.

\subsection{Host Galaxy}
\label{sec:host}

\begin{figure*}
\centerline{
\includegraphics[width=14cm]{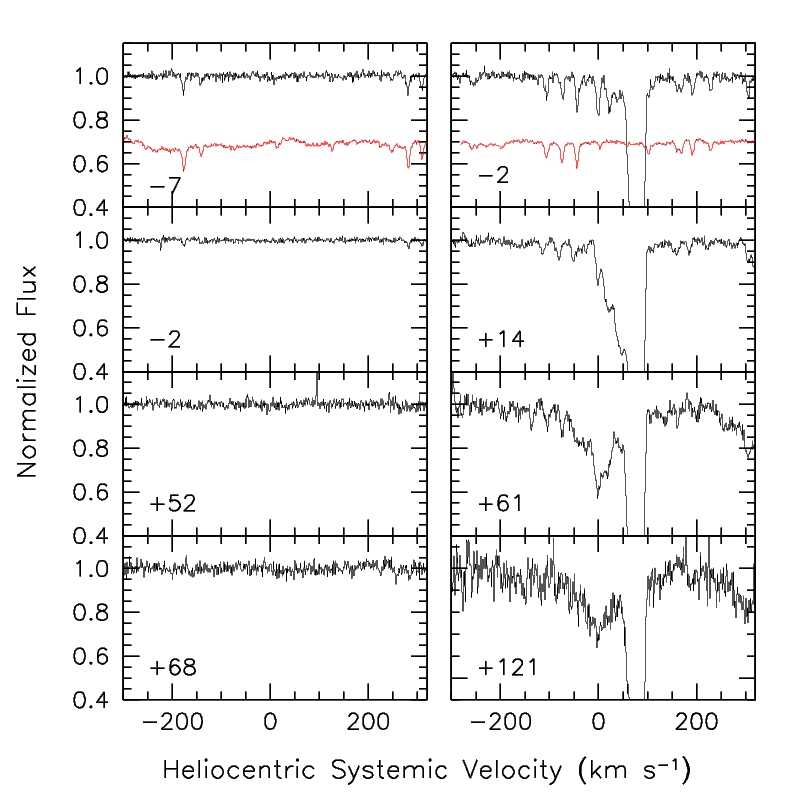}
}
\caption{\label{fig:host} The Na~I D spectral region for SN~2000cx (left)
and SN~2006X (right; P07) at four different epochs. The velocity scale
is relative to the Na~I D$_2$ component. The light-colored spectrum in
the center of the right and left top panels traces the atmospheric
absorption features in the relevant wavelength ranges.}
\end{figure*}

No trace of Na~I absorptions at velocities near the recession velocity
of the host galaxy is present in the data (Fig.~\ref{fig:host}, left
panels). All the weak absorption features visible in the spectra are
in fact identified as telluric lines, with typical $EW$s between 2 and
5 m\AA. An upper limit to the $EW$s of the Na~I D interstellar
features is $EW\lesssim$ 1 m\AA. Therefore, a very low amount of
interstellar material is expected along the line of sight within the
host galaxy. In fact, using the data for the second epoch, when the
signal-to-noise is maximum ($\sim$160), we have estimated a 3$\sigma$
upper limit for the Na~I column density
$N(Na~I)\leq$2$\times$10$^{10}$ cm$^{-2}$ for a velocity parameter
$b\leq$10 km s$^{-1}$.

\begin{figure}
\centerline{
\includegraphics[width=9cm]{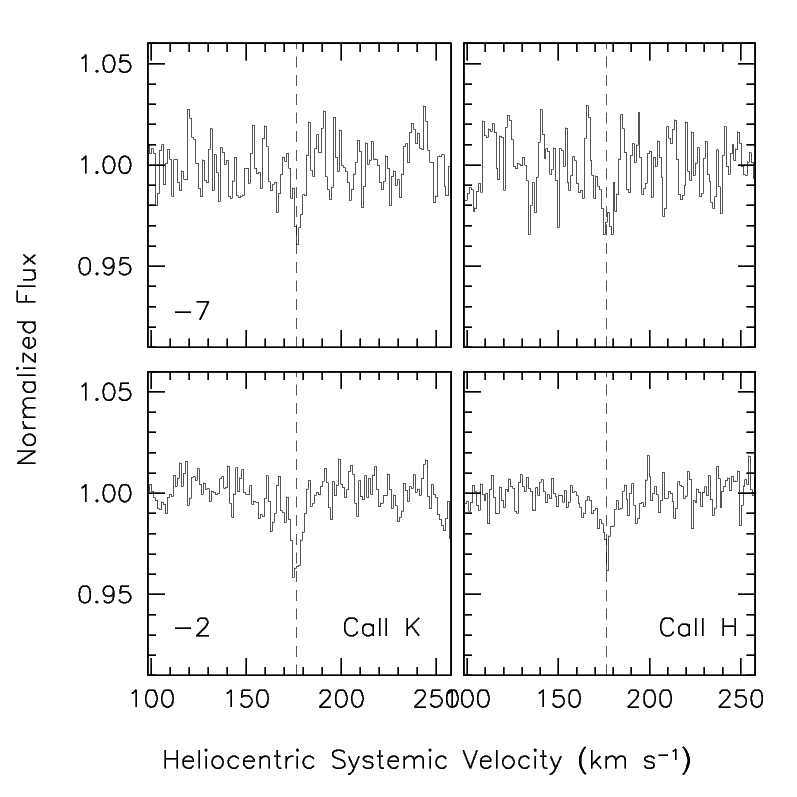}
}
\caption{\label{fig:hosthk}The Ca~II H (right) and K (left) spectral regions
for SN~2000cx on days $-$7 (top) and $-$2 (bottom). The vertical
dashed line is placed at +176.3 km s$^{-1}$.}
\end{figure}

A very weak Ca~II absorption is detected at $v$=+176.3 km s$^{-1}$ on
both $-$2 and $-$7 days (Fig.~\ref{fig:hosthk}). The column density
estimated from the K component on day $-$2 ($EW\sim$4 m\AA) is
$N(Ca~II)$=(4.4 $\pm$ 0.5) $\times$10$^{10}$ cm$^{-2}$
($b$=2.1$\pm$1.0 km s$^{-1}$). Since the upper limit for a Na~I D
component with the same velocity and velocity parameter on day $-$2 is
$N(Na~I)\leq$5$\times$10$^{9}$ cm$^{-2}$, the Na~I to Ca~II ratio is
$N(Na~I)/N(Ca~II)\leq$0.1. This is a rather low value, observed in a
few cases along the line of sight to some Galactic stars and is
usually interpreted as arising in low-density, non-molecular clouds,
where calcium depletion onto dust grains is negligible (see Crawford
\cite{crawford} and references therein).

Since the SN appears to be projected onto the approaching side of 
NGC~524 (see Sec.~\ref{sec:ngc524}), the high positive velocity
measured for the weak Ca~II absorption cannot be explained in terms of
galaxy rotation, leading us to conclude that this feature arises in a
high-velocity cloud (Wakker \& van Woerden \cite{wakker}), possibly
falling into NGC~524, and not related to the circumstellar environment
of SN~2000cx.

The absence of narrow emission lines, which would signal
interaction between the fast moving SN ejecta and the slow
circumstellar wind, was discussed by Lundqvist et
al.\ (\cite{lundqvist03}) for the two early epochs. We confirm that
there is no trace of H$\alpha$ or He~I 5876\AA\/ in the two later
epochs either.

\section{Discussion}
\label{sec:disc}

SN~2000cx exploded in the outskirts of NGC~524. While the
non-detection of interstellar lines is compatible with the peripheral
position of the SN within its host galaxy, it also marks a difference
with respect to the time-variable Na~I D features detected in the
normal Type Ia SN~2006X (Fig.~\ref{fig:host}, right panels; P07). In
SN~2006X, in fact, while the variable features were barely visible
around maximum, they were well developed two weeks later and still
clearly present two months later.  Therefore, if circumstellar gas was
located close to SN~2000cx as in SN~2006X, one might have expected no
detectable absorption at the first two epochs (days $-$7 and $-$2),
but significant absorption at the two later epochs.

The column density deduced from the most intense time-variable feature
in SN~2006X (day +14) is $N(NaI)\sim$10$^{12}$ cm$^{-2}$ (P07). If the
absorbing material is confined within a thin spherical shell with a
radius of $\sim$10$^{16}$ cm, then for solar abundance ($\log
N(Na)/N(H)$=$-$5.83; Asplund, Grevesse \& Sauval \cite{asplund}) the
implied hydrogen mass is $M(H)\sim$ 10$^{-6}$ M$_\odot$. The upper
limit set by our analysis for SN~2000cx on day +52 ($N(NaI)\leq$
5$\times$10$^{10}$ cm $^{-2}$) imposes that in this SN, everything
else being the same, $M(H)\leq$ 4$\times$10$^{-8}$ M$_\odot$
($N(H)\leq$ 3$\times$10$^{16}$ cm$^{-2}$ for a solar Na/H abundance
ratio).

The firm upper limit suggests that the circumstellar environment of
SN~2000cx is indeed very different from that of the normal Ia
SN~2006X. This can be due to a different {\it i)} progenitor system,
{\it ii)} viewing angle, {\it iii)} metallicity or {\it iv)} UV
radiation field.

\subsection{Progenitor System}

SN~2000cx has been classified as a peculiar, if not unique SN Ia (Li
et al. \cite{li}).  Therefore, one might attribute the lack of Na~I D
absorption detection to a different progenitor system, where the
circumstellar environment is significantly less dense than in the case
of the normal SN~2006X as one would predict, for instance, for the
case of a double degenerate scenario.  Following the prescriptions
given in Benetti et al.\ (\cite{benetti05}), the main
spectrophotometric parameters characterizing SN~2000cx are as
follows\footnote{The spectroscopic parameters have been derived using
unpublished spectra of SN~2006X from the Asiago-ESO-TNG supernova
archive.}: $\Delta m_{15}$(B)$_{true}$=0.94 (Li et al. \cite{li} and
assuming only Galactic reddening); $\dot{v}$=2 km s$^{-1}$\,d$^{-1}$;
$\cal R$(SiII) = 0.11; $v_{10}$(SiII) = 11.85 km s$^{-1}$1000$^{-1}$.
These parameters definitely place SN~2000cx in the low-velocity
gradient (LVG) subgroup, which includes SN~1991T and all slowly
evolving (both photometrically and kinematically) normal SNe~Ia
(Benetti et al. \cite{benetti05}). \\ A detailed analysis has shown
that this SN is most likely at the edge of normal events, bearing some
resemblance to SN~1991T (Mazzali et al. \cite{zorro}). A double
degenerate, super-Chandrasekhar scenario has been suggested for both
SN~1991T (Fisher et al. \cite{fisher}) and SN~2000cx (Li et al.
\cite{li}). In this respect, we emphasize that while SN~1991T, like
other over-luminous SNe Ia, was probably associated with a young
stellar population (Hamuy et al. \cite{hamuy96}; Howell
\cite{howell}) this is not the case for SN~2000cx, which exploded in
the outer halo of an S0 Galaxy (see Sec.~\ref{sec:ngc524} and the
discussion below).  Even though this is certainly not a stringent
argument, it suggests a different evolutionary path in the two cases,
which do display several dissimilarities (Mazzali et al.
\cite{mazzali95}; Li et al. \cite{li}).

However, the non-detection of CSM in SN~2000cx does not rule out a
more standard single-degenerate scenario. Theoretical models suggest
that there are at least two single-degenerate channels that can
produce SNe Ia, where the companion is either a main-sequence
star/subgiant (the so-called `supersoft channel') or a giant
(e.g. Hachisu, Kato \& Nomoto 1999). P07 suggested that the companion
in SN~2006X was most likely linked to the giant channel. On the other
hand, SN~2000cx may be representative of the supersoft channel (van
den Heuvel et al. \cite{vdheuvel}; Rappaport, Di Stefano \& Smith
\cite{rappaport}; Han \& Podsiadlowski \cite{han}). In this case, one
would expect a lower mass-loss rate from the system and a higher
velocity of the wind material, producing a much lower-density
circumstellar medium. Moreover, P07 speculated that the density
enhancements seen in SN~2006X could be shells of wind material that
were swept up by recurrent novae (e.g. Wood-Vasey \& Sokolosky
\cite{wood-vasey}). Without such density enhancements the Na~I D
recombination time would be very long and the absorption features
would be hard to detect.  Hence SN~2000cx and SN~2006X could easily
both have been produced through different single-degenerate
evolutionary channels (also see Parthasarathy et al. \cite{partha} and
Tutukov \& Fedorova \cite{tutukov} for recent reviews).  

In this context, Wang et al. (\cite{wangx07}), based on the
photometric and spectroscopic analysis of SN2006X, propose that this
object belongs to a subclass of Type Ia events, characterized by
distinct properties like very high expansion velocities and abnormal
extinction laws.

\subsection{CSM Geometry}

The analysis of P07 suggested that, if the progenitor of SN~2006X was
a recurrent nova, the circumstellar structure around similar SNe Ia
might only be observed for viewing angles close to the orbital
plane. If the material was concentrated in an equatorial torus, it
would be easier to decelerate the nova shells and to match the
observed velocities. Such an aspherical geometry around recurrent
novae is supported by observations of RS Ophiuchi (O'Brien et al.
\cite{obrien}; Bode et al. \cite{bode}). In SN~2006X we may have
observed the supernova through this equatorial torus, which would help
to explain the low observed velocities of the circumstellar shells
(P07). However, in general it is more likely that our line of sight
does not pass through such a torus. In the model of Bode et al.
(2007), the narrow waist covers less than a sixth of the solid angle
around the binary.  Thus, if the Na~I D lines in SN~2006X
were only seen because of a favourable inclination, this would imply
that it should only be detectable in a fraction of all SNe Ia (1 in 6
SNe in the Bode et al. model for RS Oph).

If the material observed in SN~2006X is simply the wind from a
subgiant star, it is not clear whether that stellar wind could be
aspherical enough for some of the possible lines-of-sight to avoid
passing through it.

\subsection{Metallicity}

An alternative explanation for the lack of detectable Na~I D lines in
SN~2000cx is the effect of low metallicity in the accreted material.
SN2000cx exploded in the outer regions of NGC~524, at a projected
distance of about 18 kpc from the nucleus
(Sec.~\ref{sec:ngc524}). Given the abundance gradients observed in
early-type galaxies (Carollo, Danziger \& Buson, \cite{carollo}), this
most likely implies that the SN was produced by the explosion of a low
metallicity star, as proposed by Li et al. (\cite{li}) as a possible
explanation for the observed peculiarities. In fact, a metallicity
[Na/H]=$-$1.2 would be already sufficient to make SN~2006X-like
features undetectable at the last two epochs (+52, +68). Such
sub-solar values are not rare in the outer halos of S0 galaxies (see
for instance Harris \& Harris \cite{harris}) even though it is not
clear that this is the case for NGC~524 (see Sec.~\ref{sec:ngc524}).

Even though clear relations seem to exist between the SN properties
and the parent population (Filippenko \cite{flipper}; Branch \& van
den Bergh \cite{branch94}; Hamuy et al. \cite{hamuy95}; Branch,
Romanishin \& Baron \cite{branch96}; Howell \cite{howell}), the origin
of this link seems to be age rather than metallicity (Hamuy et
al. \cite{hamuy00}; Ivanov, Hamuy \& Pinto \cite{ivanov}). Therefore,
a possible low metallicity in the outskirts of NGC~524 might be 
responsible for the absence of Na~I D circumstellar absorption but, in
the light of the currently available data, it is most unlikely that
this is the cause of the peculiarities shown by SN~2000cx (Li et
al. \cite{li}). In this context, although the models of Kobayashi et
al.\ (\cite{koba98, koba06}) predict that no SNe Ia should occur for
[Fe/H] $\lesssim$$-$1 within single-degenerate systems, some
over-luminous SNe are observed in low metallicity environments.  For
example, the 1991T-like SN~2007bk (Prieto et al. \cite{prieto2}) went
off in the outskirts of its dwarf host galaxy. Prieto et
al.\ (\cite{prieto2}) have suggested that this might be an indication
that they come from a distinct class of progenitors, such as
double-degenerate systems.

\subsection{UV deficiency}

If what we have seen in SN~2006X is a common property of normal type
Ia SNe, then the absence of time-variable absorption features in
SN~2000cx may suggest that the UV spectrum was different. In fact, in
order for the ionized Na~I to recombine on timescales of the order of
10 days, the medium has to have a sufficiently high electron density
($n_e\geq$ 10$^5$ cm$^{-3}$), which must be produced by the partial
ionization of H (P07). Since SN~2000cx was a peculiar object, one
possibility is that its UV flux was sufficient to ionize Na~I
(ionization potential 5.4 eV), but not to ionize H (13.6 eV). Under
these circumstances, the sodium recombination time would become very
large and no absorption features would appear.  However, the
ionization potential of Ca~II is rather high (11.9 eV), and under
these conditions one would reasonably expect that most calcium is in
the form of Ca~II. Therefore some H\&K absorption components should be
visible and their intensity should remain constant in time, as
observed in SN~2006X (P07). Since this is not the case, it is hard to
believe that a UV deficient SN radiation field is responsible for
the lack of Na~I lines in SN~2000cx.

\section{Conclusions}
\label{sec:concl}

Clearly, with only two objects studied it is impossible to decide why
SN~2000cx did not show the time-variable features displayed by
SN~2006X. As we have seen, the reason could be a different progenitor
channel, either a different single-degenerate channel or more
radically a double-degenerate system, an orientation effect of the CSM
or low metallicity. Many more objects need to be observed in order to
settle these open issues, requiring high-resolution spectroscopy
covering the first weeks after the explosion, when the existence of
circumstellar shells can be revealed before they are swept away by the
fast moving SN ejecta. For this purpose we have started a dedicated
campaign. The first results, obtained for SN~2007af, will be discussed
in a forthcoming paper (Simon et al. \cite{simon}).

\begin{acknowledgements}
This paper is based on observations made with ESO Telescopes at
Paranal Observatory under program IDs 65.H-0426 (P.I. P. Lundqvist)
and 65.H-0468 (P.I. M. Della Valle). The authors are indebted to
J. Liske, for his kind help during the usage of the VPGUESS package,
and to H. Kuntscher and M. Norris for providing them with the
metallicity and kinematic data for NGC~524.

\end{acknowledgements}


\begin{thebibliography}{}
\bibitem[2006]{aldering} Aldering, G., et al., 2006, ApJ, 650, 510
\bibitem[2005]{asplund} Asplund, M., Grevesse, N. \& Sauval, A.J., 2005, 
	ASPC, 336, 25
\bibitem[2000]{ballester} Ballester, P., et al., 2000, The Messenger, 101, 31
\bibitem[2005]{benetti05} Benetti, S. et al., 2005, ApJ, 623, 1011
\bibitem[2006]{benetti06} Benetti, S., et al., 2006, ApJL, 653, L129
\bibitem[2007]{bode} Bode, M.F. et al., 2007, ApJL, in press (arXiv:0706.2745)
\bibitem[1993]{branch94} Branch, D. \& van den Bergh, S., 1993, AJ, 105, 2231
\bibitem[1995]{branch} Branch, D., Livio, M., Yungelson, L.R., 
	Boffi \& F., Baron, E., 1995, PASP,  107, 1019
\bibitem[1996]{branch96} Branch, D., Romanishin, W. \& Baron, E., 1996, 
	ApJ, 465, 73
\bibitem[2004]{branch04} Branch, D., et al., 2004, ApJ, 606, 413
\bibitem[1993]{carollo} Carollo, C.M., Danziger, I.J. \& Buson, L.,
	1993, MNRAS, 265, 553
\bibitem[2000]{chornock} Chornock, R., et al., 2000, IAU Circ. 7463
\bibitem[1992]{crawford} Crawford, I., 1992, MNRAS, 259, 47
\bibitem[2000]{dekker} Dekker, H., et al., 2000, Proc. SPIE, 4008, 534
\bibitem[1991]{devauc} de Vaucouleurs, G., et al., 1991, 
	Third Reference Catalogue of Bright Galaxies, Springer Verlag, Berlin
\bibitem[2004]{emsellem} Emsellem, E., et al., 2004, MNRAS, 352, 721
\bibitem[1989]{flipper} Filippenko, A.V., 1989, PASP, 101, 588
\bibitem[1992]{flipper92} Filippenko, A.V., et al., 1992, ApJ, 384, 15
\bibitem[1999]{fisher} Fisher, A., Branch, D., Hatano, K. \& Baron, E.,
	1999, MNRAS, 304, 67
\bibitem[2004]{gerardy}Gerardy, C.L., et al., 2004, ApJ, 607, 391
\bibitem[1999]{hachisu} Hachisu, I., Kato, M. \& Nomoto, K., 1999, ApJ, 522, 487
\bibitem[2000]{hamuy00} Hamuy, M., et al., 2000, ApJ, 120, 1479
\bibitem[1995]{hamuy95} Hamuy, M., et al., 1995, AJ, 109, 1
\bibitem[1996]{hamuy96} Hamuy, M., et al., 1996, AJ, 112, 2391
\bibitem[2003]{hamuy} Hamuy, M., et al., 2003, Nature, 424, 651 
\bibitem[2004]{han} Han Z. \& Podsiadlowski Ph., 2004, MNRAS, 350, 1301
\bibitem[2002]{harris} Harris, W.E. \& Harris. G.L.H., 2002, ApJ, 123, 3108 
\bibitem[1974]{hobbs} Hobbs, L.M., 1974, ApJ, 191, 391
\bibitem[2001]{howell} Howell, D.A., 2001, ApJ, 554, L193
\bibitem[2006]{immler} Immler, S.I., et al., 2006, ApJ, 648, L119
\bibitem[2000]{ivanov} Ivanov, V., Hamuy, M. \& Pinto, P.A., 2000, ApJ, 542, 
	588
\bibitem[1998]{koba98} Kobayashi, C., Tsujimoto, T., Nomoto, K., 
	Hachisu, I. \& Kato, M., 1998, ApJL, 503, L155 
\bibitem[2000]{koba06} Kobayashi, C., Umeda, H., Nomoto, K., Tominaga, N., \&
	Ohkubo, T., 2006, ApJ, 653, 1145 
\bibitem[2006]{kunt} Kuntschner, H., et al., 2006, MNRAS, 369, 497
\bibitem[2001]{li} Li, W., et al., 2001, PASP, 113, 1178
\bibitem[2003]{lundqvist03} Lundqvist, P., et al., 2003, in 
	{\it From twilight to highlight: the physics of Supernovae}, 
	ed. W. Hillebrandt \& B. Leibundgut (Berlin: Springer), 309
\bibitem[2005]{lundqvist05} Lundqvist, P., et al., 2005, in {\it Cosmic 
 	Explosions}, ed. J. M. Marcaide, \& K. W. Weiler, CD-ROM version,
	IAU Coll., 192, 81
\bibitem[2005]{mattila} Mattila, S. et al., 2005,  A\&A, 443, 649
\bibitem[1995]{mazzali95} Mazzali, P.A., Danziger, I.J. \& Turatto, M.,
	1995, A\&A, 297, 509
\bibitem[2000]{mazzali00} Mazzali, P.A., 2000, A\&A, 363, 705
\bibitem[2005]{mazzali05} Mazzali, P.A., et al., 2005, MNRAS, 357, 200
\bibitem[2007]{zorro} Mazzali, P.A., R\"opke, F.K., Benetti, S. \& 
	Hillebrandt, W., 2007, Science, 315, 825
\bibitem[2006]{obrien} O'Brien, T.J. et al., 2006, Nature, 442, 279 
\bibitem[2006]{panagia06} Panagia, N., et al., 2006, ApJ, 469, 396
\bibitem[2007]{partha} Parthasarathy, M., Branch, D., Jeffery, D.J. \&
	Baron, E., 2007, New Astronomy Reviews, 51, 524 
\bibitem[2005]{patat05} Patat, F., Baade, D., Wang, L., Taubenberger, S. \&
	Wheeler, J.C., 2005, IAU Circ. n.~8631
\bibitem[2007]{patat} Patat, F., et al., 2007, Science, 317, 924 ({\bf P07})
\bibitem[1996]{pauldrach} Pauldrach, W.A. et al., 1996, A\&A, 312, 525
\bibitem[1999]{perlmutter} Perlmutter, S. et al., 1999, ApJ, 517, 565
\bibitem[2007]{prieto} Prieto, J.L., et al., 2007, ApJ, submitted 
	(arXiv:0706.4088)
\bibitem[2007]{prieto2} Prieto, J.L., Stanek, K.Z. \& Beacon, J.F., 2007,
	ApJ, in press, (arXiv:0706.0690)
\bibitem[2006]{quimby06} Quimby, R., et al., 2006, ApJ, 636, 400
\bibitem[1994]{rappaport} Rappaport S., Di Stefano R. \& Smith J.D., 
	1994, ApJ, 426, 692
\bibitem[1998]{riess} Riess, A.G. et al., 1998, AJ, 116, 1009
\bibitem[2002]{rudy} Rudy, R.J., et al., 2001, ApJ, 565, 413
\bibitem[1992]{lapolenta} Ruiz-Lapuente, P., et al., 1992, ApJ, 387, 33
\bibitem[2006]{sarzi} Sarzi, M., et al., 2006, MNRAS, 366, 1151
\bibitem[1998]{schlegel} Schlegel, D.J., Finkbeiner, D.P. \& Davis, M.,
	ApJ, 500, 525
\bibitem[2000]{simien} Simien, F. \& Prugniel, Ph., 2000, A\&ASS, 145, 263 
\bibitem[2007]{simon} Simon, J.D., et al., 2007, ApJ, in preparation
\bibitem[2005]{soderberg}Soderberg, A.M., 2006, Astron. Tel., 722, 1
\bibitem[2004]{sollerman04} Sollerman, J., et al., 2004, A\&A, 428, 555
\bibitem[2005]{sollerman05}Sollerman, J.,  et al., 2005, A\&A, 429, 559
\bibitem[2003]{turatto} Turatto, M., Benetti, S. \& Cappellaro, E., 2003, in
	{\it From twilight to highlight: the physics of Supernovae},
	ed. W. Hillebrandt \& B. Leibundgut (Berlin: Springer), 200 
\bibitem[2007]{tutukov} Tutukov, A.V. \& Fedorova, A.V., 2007, 
	Astronomy Reports, 51, 291
\bibitem[1992]{vdheuvel} van den Heuvel E.P.J., Bhattacharya D., 
	Nomoto K., Rappaport S., 1992, A\&A, 262, 97
\bibitem[2003]{wang03} Wang, L., et al., 2003, 591, 1110
\bibitem[2007]{wangx07} Wang, X., et al., 2007, ApJ, submitted,
	(arXiv:0708.0140)
\bibitem[1997]{wakker} Wakker, B.P. \& van Worden, H., 1997, ARAA, 35, 217
\bibitem[1973]{whelan} Whelan, J. \& Iben, I., 1973, ApJ, 186, 1007
\bibitem[2006]{wood-vasey} Wood-Vasey,W.M. \& Sokoloski, J.L., 2006, 
	ApJ, 645, L53
\bibitem[2000]{yu} Yu, C., Modjaz, M. \& Li, W.D., 2000, IAU Circ. 7458
\end{thebibliography}
\end{document}